\documentclass[aps,prb,endfloat,twocolumn,showpacs,preprintnumbers,amsmath,amssymb]{revtex4}
\usepackage{dcolumn}% Align table columns on decimal point
\usepackage[usenames]{color}
\usepackage{bm}% bold math
\usepackage{natbib}
\usepackage[dvips]{graphicx}
\usepackage{wrapfig}
\usepackage{times}
\usepackage{mathptmx}
\usepackage{textcomp}
\usepackage[usenames]{color}
\usepackage{ulem}

\begin{document}

%\preprint{APS/123-QED}

\title{Dilution effect in correlated electron system with orbital degeneracy}% Force line breaks with \\

\author{Takayoshi Tanaka,$^{\ast}$  and Sumio Ishihara}
% \altaffiliation[Also at ]{Department of Physics, Tohoku University, Sendai 980-8578, Japan}%Lines break automatically or can be forced with \\
%
% \email{takayoshi@cmpt.phys.tohoku.ac.jp}
\affiliation{%
Department of Physics, Tohoku University, Sendai 980-8578, Japan.
}%
\date{\today}% It is always \today, today,

\begin{abstract}
Theory of dilution effect in orbital ordered system is presented. The $e_g$ orbital model without spin degree of freedom and the spin-orbital coupled model in a three-dimensional simple-cubic lattice are analyzed by the Monte-Carlo simulation and the cluster expansion method. In the $e_g$ orbital model without spin degree of freedom, reduction of the orbital ordering temperature due to dilution is steeper than that in the dilute magnet. This is attributed to a modification of the orbital wave-function around vacant sites. In the spin-orbital coupled model, it is found that magnetic structure is changed from the A-type antiferromagnetic order into the ferromagnetic one. Orbital dependent exchange interaction and a  sign change of this interaction around vacant sites bring about this novel phenomena. Present results explain the recent experiments in transition-metal compounds with orbital dilution. 
\end{abstract}
\pacs{71.10.-w, 71.23.-k, 75.30.-m}
%\pacs{}% PACS, the Physics and Astronomy
\maketitle

\section{\label{sec:level1}Introduction}
Impurity effect in correlated electron system is one of the attractive themes in recent solid state physics.~\cite{imada98,maekawa_rev}  
The well known example is doping of non-magnetic impurity in high Tc superconducting cuprates; a small amount of substitution of Cu by Zn dramatically destroys the superconductivity. 
Non-magnetic impurity effect in the low-dimensional gapped spin system is another example. 
A few percent doping of Zn or Mg, which does not have a magnetic moment, into two-leg ladder systems, e.g. SrCu$_2$O$_3$, and spin-Peierls systems, e.g. CuGeO$_3$, induces long-range orders of antiferromagnetism (AFM).~\cite{azuma, sigrist96,oseroff,hase96,fukuyama96} 
Impurity effect in charge and orbital ordered state is also studied in the colossal magnetoresistive manganites.~\cite{barnabe97,kimura} 
It is reported in a so-called half-doped manganite La$_{0.5}$Ca$_{0.5}$MnO$_3$ that 
a few percent substitution of Mn by Cr collapses the charge/orbital order associated with the AFM one and induces a ferromagnetic metallic state. 
Because of no $e_g$ electrons in Cr$^{3+}$, unlike Mn$^{3+}$ with one $e_g$ electron, 
Cr is regarded as an impurity without orbital degree of freedom. 

Recently, impurity doping effect in an orbital ordered state is examined experimentally in a more ideal material. 
Murakami $\it et \ al.$ have studied substitution effect in an orbital ordered Mott insulator KCuF$_3$ with the three dimensional (3D) Perovskite crystal structure.~\cite{tatami07} 
A Cu$^{2+}$ ion in the cubic-crystalline field shows the $(t_{2g})^{6}(e_{g})^{3}$ electron configuration where one hole has the orbital degree of freedom. 
The long-range orbital order (OO), where the $d_{y^{2}-z^{2}}$- and $d_{z^{2}-x^{2}}$-like orbitals are aligned with a momentum $(\pi,\pi,\pi)$, was observed at room temperatures by several experiments. 
Since the AFM spin ordering temperature is $39$K, 
which is much lower than the OO temperature ($> 1200$K), a substitution of Cu by Zn, which has an electron configuration $(t_{2g})^6(e_g)^4$, is regarded as an orbital dilution. 
It was revealed by the resonant x-ray scattering experiments in KCu$_{1-x}$Zn$_{x}$F$_{3}$ that the OO temperature decreases with doping of Zn monotonically and the diffraction intensity 
at $(3/2\  3/2\ 3/2)$ disappears around $x=0.45$. 
At the same Zn concentration, the crystal symmetry is changed from the tetragonal to the cubic one. 
That is to say, the OO disappears around $x=0.45$. 
In dilute magnets, e.g. KMn$_{1-x}$Mg$_{x}$F$_{3}$, the $x$ dependence of the magnetic ordering temperature as well as the critical concentration where the magnetic order vanishes are well explained by the percolation theory.~\cite{elliot74,stinchcombe83}
On the contrary, the critical concentration in KCu$_{1-x}$Zn$_{x}$F$_{3}$, where the OO disappears, 
is much smaller than the site-percolation threshold in a 3D simple cubic lattice, $x_{p}=0.69$. 
These experimental observations imply that 
the dilute OO may belong to a new class of diluted systems 
beyond the conventional percolation theory. 

Dilution effect in orbital ordered state was also examined experimentally in a mother compound of the colossal magnetoresisitive manganites, LaMnO$_3$. 
The long-range OO, where the $d_{3x^{2}-r^{2}}$- and $d_{3y^{2}-r^{2}}$-like orbitals align with a momentum $(\pi, \pi, 0)$, appears below $780$K. 
The A-type AFM order, where spins are aligned ferromagnetically in the $xy$ plane and 
are antiferromagnetically along the $z$ axis, is realized at $140$K. 
Substitution of Mn$^{3+}$ by Ga$^{3+}$, which has a $3d^{10}$ electron configuration, corresponds to both the orbital and spin dilution.~\cite{farrell04,blasco02,sanches04,goodenough61,zhou01,zhou03}
From the x-ray diffraction and X-ray absorption near-edge structure (XANES) experiments, the tetragonally distorted MnO$_6$ octahedra become regular cubic ones around the Ga concentration $x=0.6$. 
That is, the OO disappear around $x=0.6$ which is smaller than the percolation threshold $x_p=0.69$ for the simple cubic lattice. 
Difference between LaMn$_{1-x}$Ga$_{x}$O$_{3}$ and KCu$_{1-x}$Zn$_{x}$F$_{3}$ is seen in the magnetic structure. 
Blasco $\it et \ al.$ observed by the neutron diffraction experiments in LaMn$_{1-x}$Ga$_{x}$O$_{3}$ 
that the ferromagnetic (FM) component appears by substitution by Ga and increases up to $x=0.5$. 
This change of the magnetic structure from the A-type AFM to FM was also confirmed by the magnetization measurements. 
This FM component cannot be attributed to the itinerant electrons 
through the double exchange interaction, since the electrical resistivity  increases with increasing $x$. 
These phenomena are in contrast to the conventional dilute magnets 
where the ordering temperature is reduced, but the magnetic structure is not changed. 
Farrell and Gehring presented a phenomenological theory for the magnetism in  LaMn$_{1-x}$Ga$_{x}$O$_{3}$.~\cite{farrell04}
They noticed that a volume in a GaO$_{6}$ octahedron is smaller than that in a MnO$_{6}$. 
Under an assumption that the Mn $3d$ orbitals around a doped Ga tend to be toward the Ga, the magnetic structure change was examined. 

In this paper, a microscopic theory of dilution effects in the $e_g$ orbital degenerate system is presented. 
We study the dilution effects in the $e_g$-orbital Hamiltonian without the spin degree of freedom, termed ${\cal H}_T$ [see Eq.~(\ref{mod2})], and the spin and $e_g$ orbital coupled one, termed ${\cal H}_{ST}$ [see Eq.~(\ref{mod1})]. 
The classical Monte-Carlo (MC) method in a finite size cluster, as well as the cluster expansion (CE) method is utilized. 
It is known that, in the classical ground state of ${\cal H}_T$ without impurity, 
a macroscopic number of orbital states are degenerated due to frustrated nature of the orbital interaction. 
We demonstrate numerically that this degeneracy is lifted at finite temperature. 
It is shown that the OO temperature decreases rapidly with increasing dilution. 
From the system size dependence of the orbital correlation function in the MC method, the OO is not realized at the impurity concentration $x=0.2$. 
The results obtained by the CE method also show rapid quenching of OO by dilution in comparison with dilute spin models. 
These results are interpreted that orbitals around impurity sites are changed so as to gain the remaining bond energy. 
This is a consequence of the bond-direction dependent interaction between the inter-site orbitals. 
In the analyses of the spin-orbital coupled model, it is shown that the A-type AFM structure realized in $x=0$ is changed into FM one by dilution. 
This is explained by changing a sign of the magnetic exchange interaction due to the orbital modification around impurity sites. 
Implications of the present microscopic theory and the experimental results in 
KCu$_{1-x}$Zn$_{x}$F$_{3}$ and LaMn$_{1-x}$Ga$_{x}$O$_{3}$ are discussed. 

In Sect.~II, the model Hamiltonian for the $e_g$ orbital degree of freedom in a cubic lattice and the spin-orbital coupled one are introduced. 
In Sect.~III, the classical MC simulation and the CE method are presented. 
Results of the numerical analyses in ${\cal H}_T$ and ${\cal H}_{ST}$ are presented in Sects.~IV and V, respectively. 
Section VI is devoted to summary and discussion. 
A part of the numerical results for the $e_g$ orbital model have been briefly presented in Ref.~\onlinecite{tanaka05}. 

\section{\label{sec:level2}Model}

Doubly degenerate $e_{g}$ orbital degree of freedom is treated by the pseudo-spin (PS) operator with magnitude of 1/2.  This operator is defined by 
\begin{equation}
\bf{T_{i}}=\frac{1}{2} \sum_{s \gamma \gamma^{\prime}} d^{\dagger}_{i\gamma s} \bf{\sigma}_{\gamma \gamma^{\prime}} d_{i\gamma^{\prime} s}, 
\end{equation}
where $d_{i\gamma s}$ is the annihilation operator of an electron with spin 
$s(=\uparrow, \downarrow)$ and orbital $\gamma(=3z^2-r^2, x^2-y^2)$ at site $i$, and $\bf{\sigma}$ are the Pauli matrices.
Occupied orbital is represented by an angle $\theta$ of PS. 
The eigen state of the $z$-component of PS with an angle $\theta$ is 
\begin{equation}
\left | \theta \right \rangle =
\mathrm{cos} \left ( \frac{\theta}{2} \right ) \left | d_{3z^{2}-r^{2}} \right \rangle 
+\mathrm{sin} \left ( \frac{\theta}{2} \right ) \left |d_{x^{2}-y^{2}} \right \rangle . 
\end{equation}
%which is the eigen state of the following operator in a rotating frame 
%\begin{equation}
%T_i^z(\theta)=\cos \theta T_i^z + \sin \theta T_i^x . 
%\end{equation}
For example, $\theta=0$, $2\pi/3$, and $4\pi/3$ correspond to the states where the $d_{3z^{2}-r^{2}}$, $d_{3y^{2}-r^{2}}$, and $d_{3x^{2}-r^{2}}$ orbitals are occupied by an electron, respectively.  
It is convenient to introduce the linear combinations of the PS operators defined by 
\begin{eqnarray}
\tau^l_i=\cos \left (\frac{2 \pi n_l}{3} \right)T_i^z-\sin \left (\frac{2\pi n_l}{3} \right) T_i^x , 
\end{eqnarray}
with $l=(x, \ y, \ z)$ and a numerical factor $(n_x, n_y, n_z)=(1, 2, 3)$. 
%
%\begin{eqnarray}
%\tau^l_i= \left \{
%\begin{array}{ll}
%\frac{1}{2} T_{iz}+\frac{\sqrt{3}}{2}T_{ix}  & l=x \\
%-\frac{1}{2}T_{iz}+\frac{\sqrt{3}}{2}T_{ix}  & l=y \\
%T_{iz} & l=z
%\end{array} \right .
%\end{eqnarray}
These are the eigen operators for the $d_{3l^2-r^2}$ orbitals. 

It is known that dominant orbital interactions in transition-metal compounds  are the electronic exchange interaction and phononic one. 
The former is derived from the generalized Hubbard-type model with the doubly degenerate $e_g$ orbitals; 
\begin{eqnarray}
{\cal H}_{\rm ele}&=&
\sum_{\langle ij \rangle \gamma \gamma' s} 
\left ( t_{ij}^{\gamma \gamma'} d_{i \gamma s}^\dagger d_{j \gamma' s}^{} +{\rm H.c.} \right )  
+U \sum_{i \gamma} n_{i \gamma \uparrow} n_{i \gamma \downarrow} 
\nonumber \\
&+&\frac{1}{2}U'\sum_{i \ \gamma \ne \gamma'} n_{i \gamma} n_{i \gamma'}
+\frac{1}{2}K \sum_{i \ \gamma \ne \gamma' s s'} 
d_{i \gamma s}^\dagger d_{i \gamma' s'}^\dagger d_{i \gamma s'}^{} d_{i \gamma' s}^{}
%\nonumber \\
%&+&I \sum_{i \ \gamma \ne \gamma'} d_{i \gamma \uparrow}^\dagger d_{i \gamma \downarrow}^\dagger 
%d_{i \gamma' \downarrow}^{} d_{i \gamma' \uparrow}^{}
,
\label{eq:hubbard}
\end{eqnarray} 
where $n_{i \gamma}=\sum_{s} n_{i \gamma s}=\sum_s d^\dagger_{i \gamma s} d_{i \gamma s}$.
We define the electron transfer integral $t_{ij}^{\gamma \gamma'}$ between the a pair of the nearest neighboring (NN) sites. 
The intra-orbital Coulomb interaction $U$, the inter-orbital one $U'$, 
and the Hund coupling $K$. 
%and the pair-hopping $I$ are introduced. 
%In an atomic limit, there is the relations $U=U'+2I$.  
%and $K=I$.   
Through the perturbational expansion with respect to the NN transfer integral under the strong Coulomb interaction, the spin-orbital superexchange model is obtained.~\cite{kugel82,ishihara97} 
By assuming a relation $U=U'+K$, for simplicity, 
it is given as 
\begin{align}
\nonumber 
{\cal H}_{\rm exc}=&-2J_{1} \sum_{\langle ij \rangle}
\left ( \frac{3}{4}+\bf{S}_{i} \cdot \bf{S}_{j} \right ) \left (\frac{1}{4}-\tau_{i}^{l}\tau_{j}^{l} \right ) \\
&-2J_{2}\sum_{\langle ij \rangle}
\left ( \frac{1}{4}-\bf{S}_{i} \cdot \bf{S}_{j} \right)
\left ( \frac{3}{4}+\tau_{i}^{l}\tau_{j}^{l}+\tau_{i}^{l}+\tau_{j}^{l} \right),
\end{align}
where $\bf{S}_{i}$ is the spin operator at site $i$ with a mgnitude of 1/2, and $l$ represents a bond direction connecting sites $i$ and $j$. 
Amplitudes of the superexchange interactions are given as 
$J_{1}[=t^{2}/(U-3K)]$ and $J_{2}(=t^{2}/U)$ 
where $t$ is the transfer integral between the $d_{3z^2-r^2}$ orbitals along the $z$ direction. 

The phononic interaction between the orbitals is derived from the orbital-lattice coupled model given by 
\begin{eqnarray}
{\cal H}_{\rm JT}&=&-g_{JT}\sum_{i m} Q_{i}^m T_{i}^m \nonumber \\
&+& \sum_{{\bf k} \xi}\frac{\omega_{{\bf k} \xi}}{2} 
\left (p^\ast_{{\bf k} \xi} p^{}_{{\bf k} \xi}+q^\ast_{{\bf k} \xi} q_{{\bf k} \xi} \right ), 
\label{eq:jt}
\end{eqnarray}
where a subscript $m$ takes $x$ and $z$. 
The first term represents the Jahn-Teller (JT) coupling 
with a coupling constant $g_{JT}$. 
Two distortion modes in a O$_6$ octahedron with the $E_g$ symmetry is denoted by $Q_{i}^z$ and $Q_{i}^x$. 
The second term is for the JT phonon where $q_{{\bf k} \xi}$ and $p_{{\bf k} \xi}$ are the phonon coordinate and momentum, respectively, and $\omega_{{\bf k} \xi}$ is the phonon frequency. 
Subscripts ${\bf k}$ and $\xi$ are the momentum and the phonon mode, respectively. 
Here, the spring constant between the NN metal and oxygen ions are taken into account. 
The interaction between orbitals and the uniform strain and the strain-energy, which are necessary in study of the cooperative JT effect, are not shown, for simplicity, in this equation. 
For convenience, the first and second terms in Eq.~\ref{eq:jt} are denoted by ${\cal H}_{\rm orb-latt}$ and ${\cal H}_{\rm latt}$, respectively. 
By introducing the canonical transformation defined by 
\begin{equation}
\widetilde q_{{\bf k} \xi}
=q_{{\bf k} \xi}-\frac{2}{\sqrt{\omega_{{\bf k} \xi}}} \sum_{m }g_{{\bf k} \xi m}^\ast T_{ -{\bf k}}^m , 
\end{equation}
and neglecting the non-commutability between ${\cal H}_{\rm lat}$ and $\widetilde q_{{\bf k} \xi}$, 
the orbital and lattice degrees of freedom are decoupled as~\cite{kanamori60,kataoka72,millis96,okamoto00} 
\begin{eqnarray}
{\cal H}_{\rm JT}=2g\sum_{ \langle ij \rangle }\tau_{i}^{l}\tau_{j}^{l} 
+{\widetilde {\cal H}}_{\rm latt}. 
\end{eqnarray}
The first term in this equation gives the inter-site orbital interaction with a coupling constant $g=g_{JT}^{2}/(3K_S)$ 
where $K_S$ is a spring constant, 
and ${\widetilde {\cal H}}_{\rm latt}$ is given by the second term in Eq.~(\ref{eq:jt}), 
i.e. ${\cal H}_{\rm latt}$, 
where the phonon coordinate and momentum are 
replaced by $\widetilde q_{{\bf k} \xi}$ and its canonical conjugate momentum $\widetilde p_{{\bf k} \xi}$, respectively. 

The model Hamiltonian studied in the present paper is given by a sum of the above two contributions. 
Quenched impurity without spin and orbital degrees of freedom 
is denoted by a parameter $\varepsilon_i$ which takes zero (one), when site $i$  is occupied (unoccupied) by an impurity. 
The Hamiltonian is given as 
\begin{align}
\nonumber 
{\cal H}_{ST}=
&-2J_{1}\sum_{\langle ij \rangle } 
\varepsilon_i \varepsilon_j \left( \frac{3}{4}+\bf{S}_{i} \cdot \bf{S}_{j} \right)
\left ( \frac{1}{4}-\tau_{i}^{l}\tau_{j}^{l} \right) \\
\nonumber 
&-2J_{2}\sum_{\langle ij \rangle} \varepsilon_i \varepsilon_j
\left ( \frac{1}{4}-\bf{S}_{i} \cdot \bf{S}_{j} \right)
\left ( \frac{3}{4}+\tau_{i}^{l}\tau_{j}^{l}+\tau_{i}^{l}+\tau_{j}^{l} \right) \\
&+2g\sum_{\langle ij \rangle }\varepsilon_i \varepsilon_j\tau_{i}^{l}\tau_{j}^{l} . 
\label{mod1}
\end{align}
Numerical results in this Hamiltonian is presented in Sect.~\ref{sec:level5}.   
We also study dilution effect in the orbital model without spin degree of freedom. 
This model is given by taking $\bf{S}_{i} \cdot \bf{S}_{j}$ in Eq.~(\ref{mod1}) to be zero. 
This procedure may be justified in the diluted orbital system of KCu$_{1-x}$Zn$_{x}$F$_{3}$ where the N$\rm \acute e$el temperature (T$_{\rm N}$) is much below the OO temperature $T_{\rm OO}$. 
The explicit form of the $e_g$ orbital model without spin degree of freedom is given by 
\begin{align}
{\cal H}_{T}=2J\sum_{\langle ij \rangle}\varepsilon_i\varepsilon_j \tau_{i}^{l}\tau_{j}^{l} , 
\label{mod2}
\end{align}
where $J(=2g+3J_{1}/4-J_{2}/4)$ is the effective coupling constant.
Numerical results of this model Hamiltonian are presented in Sect.~\ref{sec:level4}.

\section{\label{sec:level3} Method}

In order to analyze the model Hamiltonian introduced above by using the unbiased method, 
we adopt mainly the classical MC simulation in finite size clusters. 
The orbital PS operator is treated as a classical vector defined in the $T_{z}-T_{x}$ plane, 
i.e. $T_i^z=(1/2)\cos \theta_i$ and $T_i^x=(1/2)\sin \theta_{i}$ where $\theta_{i}$ is 
a continuous variable.
As well as the conventional Metropolis algorithm, 
the Wang-Landau (WL) method is utilized.~\cite{wang01}
This is suitable for the present spin-orbital coupled model where the energy scales of 
the two degrees  are much different with each other. 
In order to calculate the density of state, $g(E)$, with high accuracy in the WL method, 
we take that the minimum energy edge $E_{\rm min}$ in $g(E)$ is higher a little than the ground state 
energy $E_{\rm GS}$, and assume $g(E_{\rm GS}<E<E_{\rm min})=0$. 
As a result, the present MC simulation is valid above a characteristic temperature $T_{\rm min}$ 
which is determined by $|E_{\rm min}-E_{\rm GS}|$. 
This situation will be discussed in Sect.~\ref{sec:level4} in more detail. 
%\textcolor{blue}{
%In order to calculate a continuous systems by WL method, it is needed that the energy is divided into bins. 
%And the minimum energy (E$_{min}$) need be set up for calculating the density of states (DOS) $g(E)$.
%E$_{min}$ is larger than the ground state energy(E$_{0}$), and the DOS lower than E$_{min}$ is set to be zero.
%Any physical quantity $A$ is calculated by WL method is expressed the formula $<A>=(\sum_{E}A(E)g(E)e^{-E/T})/(\sum_{E}g%(E)e^{-E/T})$ 
%where $E$ represents the energy, and $\sum_{E}$ represents a sum of bins,
%and $A(E)$ is Monte Carlo average of $A$ at $E$.
%Hence, the temperature dependence disappears in low temperature range.
%But, if E$_{min}$ takes sufficiently low, magnetic order and OO can be discussed.
%}
The simulations have been performed in $L \times L \times L$ cubic lattices ($L=12 \sim 18$) with the periodic-boundary condition.
In the Metropolis method, for each sample, $3 \times 10^4-1 \times 10^5$MC steps are spent for measurement after $8 \times 10^3-2 \times 10^4$ MC steps for thermalization. 
Physical quantities are averaged over $20-80$ samples at each parameter set.
In the WL method, 
the final modification factor~\cite{wang01} is set to be $f_{final}=\exp(2^{-27})$. 
After calculating the density of states, 
2$\times$10$^{7}$ MC steps are spent for measurement.

To supplement the classical MC simulation, the ordering temperatures are also calculated by utilizing the CE method. 
We apply the CE method proposed in Ref.~\onlinecite{mano77} to the present orbital model. 
For a given impurities configuration $\{ \varepsilon \}$ in a lattice with $N$ sites, 
the OO parameter is given as
\begin{align}
M_{\{ \varepsilon \}}=\mathrm{Tr}_N  \sum_{i} \varepsilon_iT_{i}^z \rho_{N \{ \varepsilon \}} ,
\label{met1}
\end{align}
with the density matrix  
\begin{equation}
\rho_{N \{ \varepsilon \}}=\frac{e^{-\beta {\cal H}_{\{ \varepsilon \}} }}{\mathrm{Tr}_N 
e^{-\beta {\cal H}_{\{ \varepsilon \}} }} , 
\end{equation} 
where ${\rm Tr}_N$ represents the trace over the PS operator at sites with $\varepsilon_i=1$ 
in a crystal lattice, 
and ${\cal H}_{\{ \varepsilon \}}$ is the Hamiltonian with impurity configuration $\{ \varepsilon \}$. 
The OO parameter per site is obtained by averaging about all 
possible impurity configuration $\{ \varepsilon \}$ as 
\begin{equation}
M= \frac{1}{(1-x)N} \left \langle M_{\{ \varepsilon \}} \right \rangle_{\{ \varepsilon \}} , 
\label{eq:mmm}
\end{equation}
where $x$ is the impurity concentration. 
In the CE method, a cluster consisting of $m$ sites, termed $\{ m \}$, is considered, 
and the PS operators which do not belong to $\{m \}$ are replaced by stochastic variables $\sigma_i$. 
Here we take $(T_i^x, T_i^z)=(0, \sigma_{i})$.
The effective Hamiltonian thus obtained is denoted as ${\cal H}_{\{\varepsilon \} \{ m \} \{ \sigma \}}$ where $\{\sigma \}$ is a set of $\sigma_i$, and the corresponding density matrix is 
\begin{align}
\rho_{\{\varepsilon \} \{ m \} \{\sigma\}}=
\frac{\exp \left (-\beta {\cal H}_{\{ \varepsilon \} \{ m \} \{\sigma\}} \right)}
{\mathrm{Tr}_{\{ m \}} \exp \left( -\beta {\cal H}_{\{ \varepsilon \} \{ m \} \{\sigma\}} \right )} , 
\label{met2}
\end{align}
where ${\rm Tr}_{\{ m \}}$ represents the trace over the PS operators in a cluster $\{m \}$. 
We expand $M_{\{\varepsilon \}}$ into a series of cluster averages as follows, 
\begin{align}
\nonumber M_{\{ \varepsilon \}}&=\sum_{m=1}^{N}\sum_{\{m\}}\sum_{k=1}^{m}\sum_{\{ k \}}(-1)^{k-m} 
\nonumber \\
&\times \mathrm{Tr}_{\{ k \}} 
\left[ \left ( \sum_{i \in \{ k \}} \varepsilon_{i}T_i^z \right) 
\sum_{\{\sigma\}}\rho_{\{ \varepsilon \} \{k \} \{\sigma\} } \right],
\label{met3}
\end{align}
where  $\sum_{\{m\}}$ is taken over all possible clusters consisting of $m$ sites, and $\sum_{\{k\}}$ is taken over all subclusters of $k$ sites belonging to a given $\{m\}$. 
%$\sum_{\{\sigma\}}$ represents a sum over $\{\sigma\}$ with probability $P(\sigma_{i})$.
%
The variable $\sigma_{i}$ takes $1/2$ or $-1/2$ by a probability of 
\begin{eqnarray}
P(\sigma_{i})=\delta_{\sigma_{i},\frac{1}{2}} 
\left (\frac{1+2 M }{2} \right ) +\delta_{\sigma_{i},-\frac{1}{2}} 
\left ( \frac{ 1-2 M}{2} \right ). 
\label{eq:prob}
\end{eqnarray}
By solving Eqs.~(\ref{eq:mmm})-(\ref{eq:prob}) self-consistently, the order parameter and the ordering temperature are obtained as a function of impurity concentration. 
In the present study, we adopt the CE method in the two-site cluster approximation, i.e. $m=2$. 
It was shown that, even in the two-site cluster approximation, 
the obtained results show good accuracy in the case of the ferromagnetic Heisenberg model in a simple cubic lattice; 
deviations from the results by other reliable methods are about $2\%$ for the critical impurity concentration.~\cite{mano77}
To compare the results in the classical MC simulation, 
the ordering temperature is also calculated in the classical version of the CE method where the traces in Eqs.~(\ref{met2}) and (\ref{met3}) are replaced by integrals with respect to the continuous variable $T_{i}^z$ between $1/2$ and $-1/2$. 

\section{\label{sec:level4} Dilution in the $e_{g}$ orbital model}

\begin{figure}
\begin{center}
\includegraphics[width=\columnwidth]{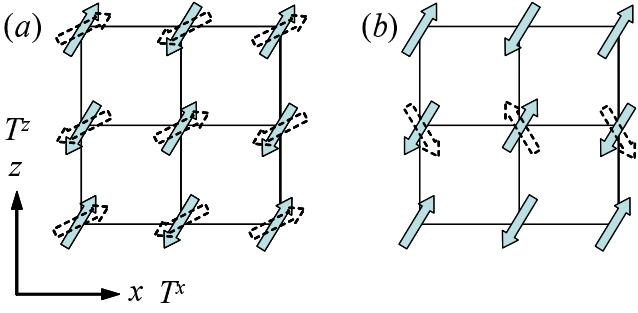}
%\scalebox{1.0}{\includegraphics[width=5.5cm,height=4.5cm,clip]{egspesificsize.eps}}
\end{center}
\caption{(a) Scematic picture of the degenerate PS configurations termed the type-(I) degeneracy, and (b) that of the type-(II) one.}
\label{fig:mfdege}
\end{figure}
In this section, numerical results for dilution effects in the $e_g$ orbital Hamiltonian (\ref{mod2}) are presented.
First, we show the results in the MC simulation without impurity.
It is known that there is a macroscopic degeneracy in the mean-field (MF) ground state in ${\cal H}_T$ without impurity.~\cite{nussinov04}
This degeneracy is classified into the following two types: 
(I) Consider a staggered-type OO with two sublattices, 
termed A and B, and momentum $\bf Q=(\pi,\pi,\pi)$. 
In the MF ground state, the PS angles in the sublattices are given by ($\theta_{A},\theta_{B}$)=($\theta,\theta+\pi$) with any value of $\theta$. Such continuous rotational symmetry is unexpected from the Hamiltonian ${\cal H}_{T}$ where any continuous symmetries do not exist.
(II) Consider an OO with $\bf Q=(\pi,\pi,\pi)$ and ($\theta_{A},\theta_{B}$)=($\theta_{0},\theta_{0}+\pi$), and focus on one direction in three-dimensional simple-cubic lattice, e.g., the $z$ direction. 
The MF energy is preserved by changing all PS in each layer perpendicular to the $z$ axis independently as ($\theta_{0},\theta_{0}+\pi$) $\to$ ($-\theta_{0},-\theta_{0}-\pi$). 
These are schematically shown in Fig.~\ref{fig:mfdege}. 
Both types of degeneracy are understood from the momentum representation of the 
orbital interaction, 
\begin{equation} 
{\cal H}_{T}=2J\sum_{\bf{k}} \psi^\dagger_{\bf{k}}\hat{E}(\bf{k})\psi_{\bf{k}} , 
\end{equation}
with 
$\psi_{k}=[T_{\bf{k}}^z, T_{\bf{k}}^x]$ and the $2 \times 2$ matrix $\hat{E}(\bf{k})$.
By diagonalizing $\hat{E}(\bf{k})$, we obtain the eigen values 
\begin{eqnarray}
E_{\pm}(\bf{k})&=&c_{x}+c_{y}+c_{z} 
\nonumber \\
&\pm& \sqrt{c_{x}^{2}+c_{y}^{2}+c_{z}^{2} -c_{x}c_{y}-c_{y}c_{z}-c_{z}c_{x}} , 
\end{eqnarray}
where $c_{l}=\cos a k_{l}$ with a lattice constant $a$.
The lower eigen value $J_{-}(\bf{k})$ has its minima along $(\pi,\pi,\pi)-(0,\pi,\pi)$ and other two-equivalent directions.~\cite{ishihara_ol} 
At the point $\Gamma$, the two eigen values $E_+({\bf k})$ and $E_-({\bf k})$ are degenerate. 
That is, the orbital states corresponding to these momenta are energetically degenerate in the MF level.  
A lifting of this degeneracy in the MF ground state has been examined  from the view points of the order-by-disorder mechanism by utilizing the spin wave analyses.~\cite{nussinov04,brink99,kubo02}

%\begin{figure}
%\begin{center}
%\includegraphics[width=0.8\columnwidth]{specific.eps}
%\scalebox{1.0}{\includegraphics[width=5.5cm,height=4.5cm,clip]{egspesificsize.eps}}
%\end{center}
%\caption{Temperature dependence of specific heat at $x=0$.}
%\label{egspecific}
%\end{figure}
%
\begin{figure}
\begin{center}
\includegraphics[width=0.8\columnwidth]{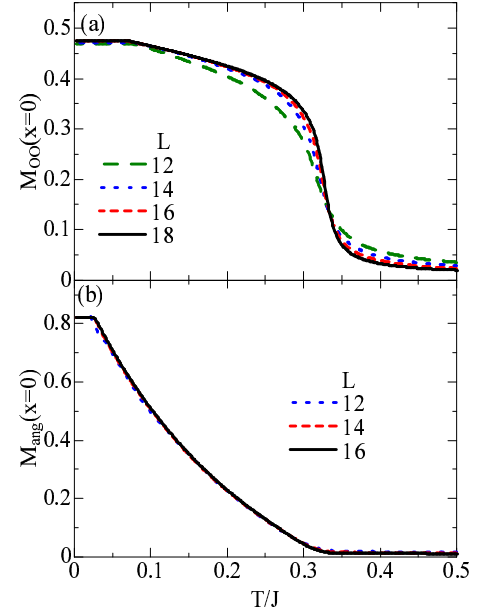}
%\scalebox{1.0}{\includegraphics[width=9cm,height=4.5cm,clip]{moo.eps}}
\end{center}
\caption{(a) System size dependence of the orbital correlation function $M_{\rm OO}(x=0)$, 
and (b) that of the orbital angle function $M_{\rm ang}(x=0)$. 
The minimum energy $E_{\rm min}$ in the WL method is taken to be $0.95E_{\rm GS}$ in (a) and $0.98E_{\rm GS}$ in (b). 
}
\label{egorder}
\end{figure}
Here we demonstrate the degeneracy lifting and appearance of the long-rage OO 
by the MC method.
%
%Temperetura dependence of the specific heat is shown in Fig.\ref{egspecific} 
%which takes its maximum value around $T/J=0.32$ in the case of $L=12$.
%With increasing the system size, 
%a peak structure becomes sharp and a peak position gradually increases. 
%
We introduce, for impurity concentration $x$, the staggered orbital correlation function 
\begin{equation}
 M_{\rm OO}(x)=\frac{1}{N(1-x)}
%\sqrt{
\left \langle \biggl \{ \sum_{i} (-1)^{i} \varepsilon_i {\bf T}_{i} \biggr \}^{2} \right \rangle^{1/2} , 
%}, 
\label{eq:moo}
\end{equation}
and the angle correlation function 
\begin{equation}
M_{\rm ang}(x)=\frac{1}{N(1-x)} 
%\sqrt{ 
\left \langle  \biggl \{ \sum_{i} (-1)^{i} 
\varepsilon_i \cos 3\theta_{i} \biggr \}^{2} \right \rangle^{1/2}  , 
%},
\end{equation}
where $\langle \dots \rangle$ represents the MC average and $N=L^{3}$.
The orbital correlation at the momentum ${\bf Q}=(\pi, \pi, \pi)$ is represented by $M_{\rm OO}(x)$, 
and the angle correlation $M_{\rm ang}(x)$ takes one, when the orbital PS angle is $2\pi n/3$ with an integer number $n$. 
Therefore, $M_{\rm OO}(x)$ and $M_{\rm ang}(x)$ 
are utilized as monitors for lifting of the type-(II) and (I) degeneracies, respectively.  
Temperature dependences of $M_{\rm OO}(x=0)$ for various $L$ are shown in Fig.~\ref{egorder}(a). 
With decreasing temperature, calculated results for all $L$ show a sharp increasing around $T/J=0.35$.
This increasing becomes sharper with the system size $L$. 
Below $T/J=0.08$, $M_{\rm OO}(x=0)$ takes a temperature-independent value of about $0.47$. 
This flat behavior is attributed to the lowest energy edge $E_{\rm min}$ for the density of state calculated in the WL method, as explained in Sect.~\ref{sec:level3}. 
An extrapolated value of $M_{\rm OO}(x=0)$ toward $T=0$ is close to 0.5 which indicates that the type-(II) degeneracy is lifted and the OO with the momentum ${\bf Q}=(\pi, \pi, \pi)$ is realized.  
Temperature dependences of $M_{\rm ang}(x=0)$ presented in Fig.~\ref{egorder}(b) increase monotonically toward one in the low temperature limit. 
Almost no-size dependence is seen in $M_{\rm ang}(x=0)$. 
Therefore, the type-(I) degeneracy is also lifted and the PS angle is fixed.  
Both results indicate the long-range OO where the momentum is ${\bf Q}=(\pi, \pi, \pi)$, and the PS angles are $(\theta_A, \theta_B)=(\theta_0, \theta_0+\pi)$ with $\theta_0=2\pi n/3$.

\begin{figure}
\begin{center}
\includegraphics[width=0.8\columnwidth]{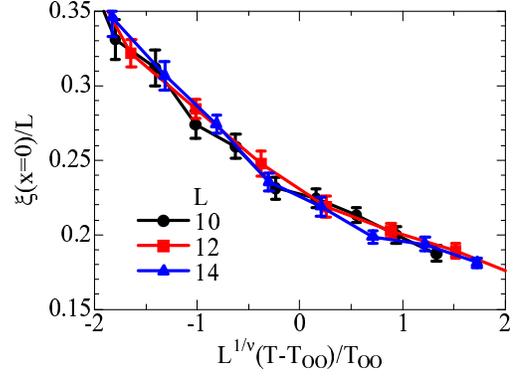}
%\scalebox{1.0}{\includegraphics[width=5.5cm,height=4.5cm,clip]{egscalingx0.eps}}
\end{center}
\caption{Scaling plot of the correlation length $\xi(x=0)$ for the staggered orbital correlation.
Numerical data are obtained by the Metropolis algorithm.}
\label{egcorrelation}
\end{figure}

The temperature at which $M_{\rm OO}(x=0)$ and $M_{\rm ang}(x=0)$ change abruptly is around $T/J=0.33$ corresponding to the OO temperature $T_{\rm OO}(x=0)$. 
In more detail, this temperature is determined by the finite-size scaling for the correlation length.
This is calculated by the second-moment method;
\begin{equation}
\xi(x)=\frac{1}{2\sin (ak_{\rm min}/2) } 
\sqrt{ 
\frac{ M_{\rm OO}(x)^{2}-  M_{k_{\rm min}}(x)^{2}  }
{ M_{k_{\rm min}}(x)^{2}  } 
}, 
\end{equation}
with 
\begin{equation}
M_{k_{\rm min}}(x)=\frac{1}{N(1-x)} 
\left \langle \left \{ 
\sum_{i} e^{i (\bf{Q-k}) \cdot \bf{r_{i}}} \varepsilon_i \bf{T}_{i}   
\right \}^2 \right \rangle^{1/2}
, 
\label{eg2}
\end{equation}
where $k_{\rm min}=(2\pi/L, 0, 0)$.
%The correlation lengths are calculated by the Metropolis method.
%Accoding to the scaling theory, 
The scaling relation for $\xi(x)$ is 
\begin{align}
\xi(x)=L F \left [ L^{1/\nu} \left \{ T-T_{\rm OO}(x) \right \} \right ],
\label{eg3}
\end{align}
where $\nu$ is the critical exponent for correlation length, and 
$F$ is the scaling function.
The correlation lengths $\xi(x=0)/L$ for various sizes cross with each other at $T_{\rm OO}(x=0)$. 
In Fig.~\ref{egcorrelation}, we plot $\xi(x=0)/L$ as a function of $L^{1/\nu}[T-T_{\rm OO}(x=0)]$. 
The scaling analyses work quite well for $L=10$, 12, and 14.
The OO temperature $T_{\rm OO}(x=0)$ and the critical exponent $\nu$ are determined by the least-square fitting for the polynomial expansion. 
We obtain as $T_{\rm OO}(x=0)/J=0.344 \pm 0.002$ and $\nu=0.69-0.81$, although statistical errors are not enough to obtain the precise value of $\nu$.

\begin{figure}
\begin{center}
\includegraphics[width=0.8\columnwidth]{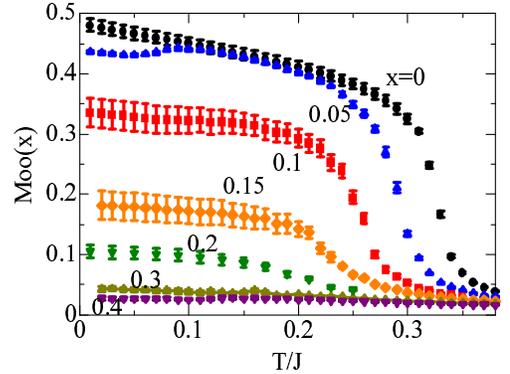}
%\scalebox{1.0}{\includegraphics[width=5.5cm,height=4.5cm,clip]{egorderimpurities.eps}}
\end{center}
\caption{Impurity concentration dependence of the staggered orbital correlation function $M_{\rm OO}(x)$. 
System size is taken to be $L=18$. 
Numerical data are obtained by the Metropolis algorithm.}
\label{egorderimpurity}
\end{figure}
\begin{figure}
%\begin{center}
\includegraphics[width=0.8\columnwidth]{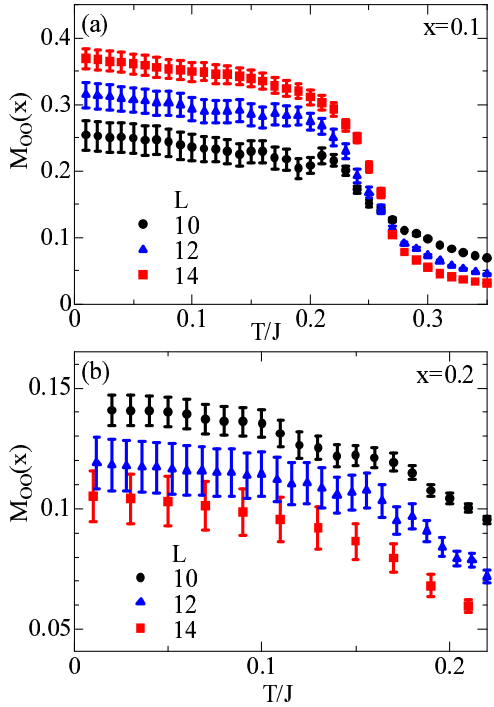}
%\scalebox{1.0}{\includegraphics[width=9cm,height=4.5cm,clip]{egordercomparison.eps}}
%\end{center}
\caption{(a) System size dependence of the orbital correlation function $M_{\rm OO}(x)$ at $x=0.1$, and (b) that at $x=0.2$.}
\label{egordercompar}
\end{figure}
Now, we examine impurity effect in the OO.
In Fig.~\ref{egorderimpurity}, we present the staggered orbital correlation function 
$M_{\rm OO}(x)$ for several impurity concentration $x$. 
Numerical data are obtained by the Metropolis algorithm in the classical MC method and the system size is chosen to be $L=18$.
First, we focus on the region of $x \leq 0.15$. 
As shown above, $M_{\rm OO}(x=0)$ abruptly increases at $T_{\rm OO}(x=0) \sim 0.34J$ and is saturated to $0.5$ in the low temperature limit. 
By introducing impurity, $M_{\rm OO}(x > 0)$ does not reach $0.5$ even at $T/J=0.01$, and its saturated value in low temperatures gradually decreases with increasing $x$. 
Although the system sizes are not sufficient to estimate $M_{\rm OO}(x)$ in the thermodynamic limit, $M_{\rm OO}(x > 0)$ at zero temperature does not show the smooth convergence to 0.5 in contrast to the diluted spin models.
Beyond $x=0.15$, results are different qualitatively; although $M_{\rm OO}(x)$ starts to increase around a certain temperature (e.g. $T/J \sim 0.24$ at $x=0.2$), saturated values of $M_{\rm OO}(x)$ in the low temperature limit are rather small. 
In order to compare the size dependences of $M_{\rm OO}(x)$, temperature dependences of $M_{\rm OO}(x=0.1)$ and $M_{\rm OO}(x=0.2)$ for several system sizes are presented 
in Fig.~\ref{egordercompar}.
In Fig.~\ref{egordercompar}(a) for $x=0.1$, $M_{\rm OO}(x)$ for several sizes cross around $T/J=0.25$ below which 
$M_{\rm OO}(x)$ increases with $L$. 
On the other hand, In Fig.~\ref{egordercompar}(b) for $x=0.2$, 
$M_{\rm OO}(x)$ monotonically decreases with $L$ in all temperature range.
\begin{figure}
\begin{center}
\includegraphics[width=0.8\columnwidth]{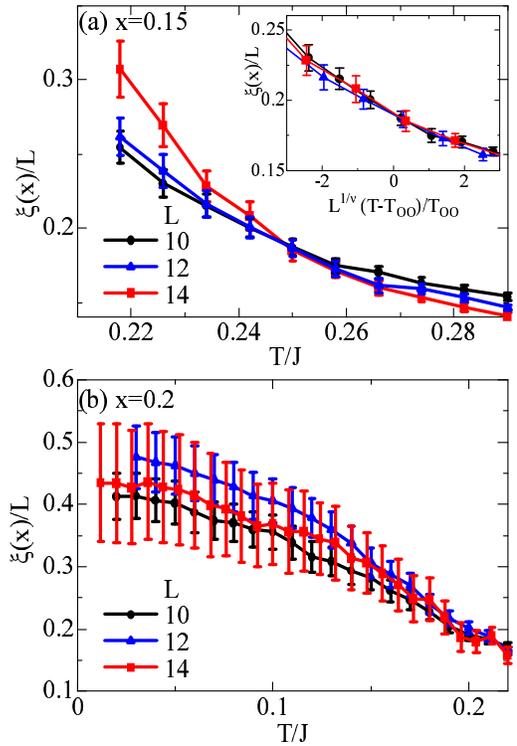}
%\scalebox{1.0}{\includegraphics[width=9cm,height=4.5cm,clip]{egcorrelationcomparison.eps}}
\end{center}
\caption{(a) System size dependence of the correlation length $\xi(x)/L$ at $x$=0.15, and 
(b) that at $x$=0.2. The inset of (a) is the scaling plot for $\xi(x)$ at $x=0.1$. 
The OO temperature and the critical exponent at $x=0.15$ are obtained to be 
$T_{\rm OO}(x)=0.248\pm0.003$ and $\nu=0.755 \pm 0.085$, respectively.}
\label{egcorrelationcompar}
\end{figure}
This difference above and below $x=0.15$ is also seen in the results of the correlation length. 
In Fig.~\ref{egcorrelationcompar}, a correlation length at $x=0.15$ and $x=0.2$ are compared.
In $x=0.15$, $\xi(x)$ for different sizes cross around $T/J=0.25$. 
As shown in the inset of Fig.~\ref{egcorrelationcompar}(a), the scaling analyses works well. 
From this analyses for $\xi(x)$, the OO temperature in $x=0.15$ 
is obtained as $T_{\rm OO}(x=0.15)/J=0.248 \pm 0.003$.
On the other hand, in $x=0.2$ [see Fig.~\ref{egcorrelationcompar}(b)], 
$\xi(x)$ for different sizes do not seem to cross with each other at a certain temperature, and the scaling analyses does not work.
From the above numerical results, it is thought that the long-range OO disappears around $0.15 < x <0.2$.     

\begin{figure}
\begin{center}
\includegraphics[width=0.8\columnwidth]{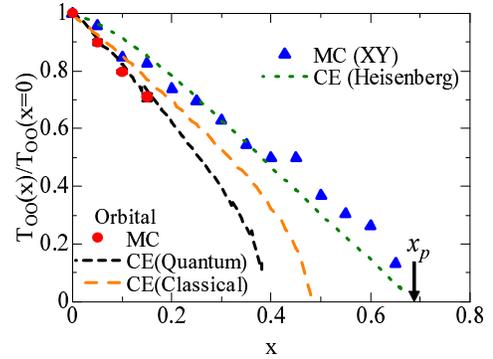}
%\scalebox{1.0}{\includegraphics[width=6.5cm,height=5cm,clip]{egphasecomphei.eps}}
\end{center}
\caption{Impurity concentration $x$ dependence of the OO temperature $T_{\rm OO}(x)$. 
Filled circles are obtained by the MC method. 
Results by the quantum  and classical CE method are shown by broken lines.
For comparison, $x$ dependence of the N$\rm \acute e$el temperature $T_{\rm N}(x)$ in the 3D XY model obtained by the MC method and that in 3D Heisenberg model by the classical CE one are presented by filled triangles and dotted line, respectively. 
Thick arrow indicates the percolation threshold in a 3D simple cubic lattice. 
}
\label{egphase}
\end{figure}
The impurity concentration $x$ dependence of $T_{\rm OO}(x)$ obtained by the MC and CE methods 
are presented in Fig.~\ref{egphase}. 
Two kinds of the CE methods, where the PS operators are treated as classical vectors and quantum operators, are carried out. 
These are termed the classical and quantum CE methods, respectively. 
In both cases, we adopt the two-size cluster. 
As a comparison, the N${\rm \acute e}$el temperatures in the 3D XY model obtained by the classical MC method, 
and those in the 3D Heisenberg model by the classical CE method are also plotted in the same figure.  
It is shown that decrease of $T_{\rm OO}(x)$ by the MC method is much steeper than that of $T_{\rm N}(x)$ in the XY and Heisenberg models. 
As shown in the size dependences of $M_{\rm OO}(x)$ and $\xi(x)$ at $x=0.2$, it is thought that the long range OO is not realized at this impurity concentration. 
A rapid decrease of $T_{\rm OO}(x)$ in comparison with the spin ordering temperatures is also obtained by the CE method. 
The OO temperature monotonically decreases with $x$, and disappears around $x=0.4$ in the quantum CE calculation, and around $0.5$ in the classical CE one. 
The critical impurity concentrations obtained by the MC and CE methods are much smaller than the percolation threshold 
$x_p=0.69$ in the 3D simple-cubic lattice.

\begin{figure}
\begin{center}
\includegraphics[width=0.6\columnwidth]{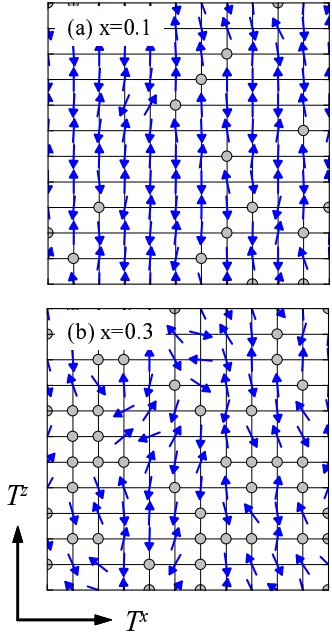}
%\scalebox{1.0}{\includegraphics[width=8cm,height=4.5cm,clip]{egPsconfigulation.eps}}
\end{center}
\caption{(a) A snapshot in the MC simulation for the PS configuration at $x$=0.1, and (b) that at $x=0.3$. 
Filled circles indicate impurities.}
\label{PScofig}
\end{figure}
Let us explain the physical picture of the orbital dilution. 
Snapshots of the PS configuration in the MC simulation are shown in Figs.~\ref{PScofig}(a) and (b) for $x=0.1$ and 0.3, respectively.
The staggered-type OO with the orbital angle ($\theta_{A}, \theta_{B}$)=(0,$\pi$) is seen in the background of 
Fig.~\ref{PScofig}(a).
At the neighboring sites of the impurities indicated by the open circles, PS vectors tilt from the angle of $(0, \pi)$.
This deviation of the PS angles is not only due to the thermal fluctuation. 
Focus on the NN sites along the $x$ direction of an impurity which occupies the down PS sublattice. 
In almost all these sites, 
PS angles are changed from $0$ to a positive angle $\delta \theta$. 
This kind of tilting from $(0, \pi)$ becomes remarkable at $x=0.3$. 
\begin{figure}
\begin{center}
\includegraphics[width=\columnwidth]{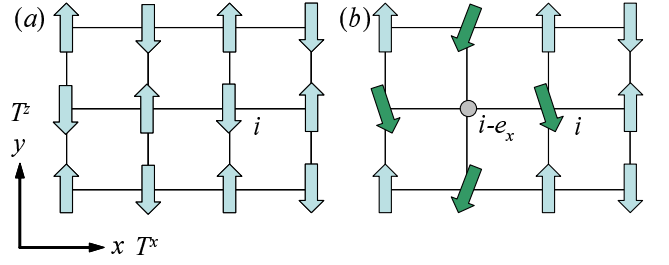}
%\scalebox{1.0}{\includegraphics[width=8cm,height=5cm,clip]{egPSmeanF.eps}}
\end{center}
\caption{(a) A schematic PS configuration without impurity, and (b) that with an impurity.
A filled circle represents an impurity. 
}
\label{PSdisturb}
\end{figure} 
Then, we explain the microscopic mechanism of this PS tilting due to dilution [see Fig.~\ref{PSdisturb}]. 
Focus on a PS at a certain site termed $i$. 
The interaction acting on this site is considered by the MF approximation where 
we assume the staggered-type OO with the PS angle $(0, \pi)$ except for the site $i$ and an impurity site. 
The Hamiltonian which concerns the interaction acting on this site is given as 
\begin{eqnarray}
{\cal H}_{T}^{(i)}&=&2J  \sum_{l=(x, y, z)}
\left \langle 
\varepsilon_{i+\hat{e}_l} \tau^{l}_{i+\hat e_l}+\varepsilon_{i-\hat{e}_l}\tau^{l}_{i-\hat{e}_l} 
\right \rangle  \tau^{l}_{i}  \nonumber \\
&=&- \sum_{l=(x, y, z)}  \bf h_l \cdot \bf T_i , 
\label{eg5}
\end{eqnarray}
where $\hat{e}_l$ is a unit vector along $l$ in the simple cubic lattice, 
and ${\bf h}_l=(h_l^x, h_l^z)$ are the MF.
In the case of no dilution [Fig.~\ref{PSdisturb}(a)], 
the mean-fields are given by 
${\bf h}_x=J(-\sqrt{3}/2, -1/2)$, ${\bf h}_y=J(\sqrt{3}/2, -1/2)$
${\bf h}_z=J( 0, -2)$, and the Hamiltonian in Eq.~(\ref{eg5}) is reduced to 
\begin{align}
{\cal H}_T^{(i)}=3JT_{i}^z . 
\label{eg7}
\end{align}
This implies that the stable PS configuration at the site $i$ is $\theta_i=\pi$. 
Then, introduce an impurity at site $i-\hat e_x$ and consider the PS at site $i$ [Fig.~\ref{PSdisturb}(b)]. 
The $x$-component of the MF in the case without impurity 
is changed into 
${\bf h}_x=J(-\sqrt{3}/4, -1/4)$, and others are not. 
The effective interaction in Eq.~(\ref{eg5}) is given as  
\begin{equation}
{\cal H}_{T}^{(i)}=\frac{J}{4} \left ( 11 T_{i}^{z}-\sqrt{3}T_{i}^{x} \right ) , 
\label{eg8}
\end{equation}
implying that the stable orbital angle at site $i$ is $\theta_i \sim \pi-0.15$.
This PS tilting due to dilution is attributed to the fact that the orbital interaction explicitly depends on the bond direction and is the essence of the diluted orbital systems. 
This is highly in contrast to the dilute spin system where dilution does not cause specific spin tilting around the impurity site but simply increases thermal spin fluctuation since number of the interacting bond is reduced.

\section{\label{sec:level5} Dilution in the spin-orbital model}

In this section, we examine the dilution effect in the spin-orbital coupled model described by ${\cal H}_{ST}$ in Eq.~(\ref{mod1}). 
First, we briefly introduce the MF calculation for the spin and orbital structures at $x=0$.
The two sublattice structures for both the spin and orbital ordered states are considered, and the PS angles in sublattices A and B are assumed to be 
$(\theta_{\rm A}, \theta_{\rm B})=(\theta, -\theta)$. 
We obtain the ferromagnetic spin order in the case of $J_{1}/J_{2} \geq 3$, 
and the A-type AFM one in $J_{1}/J_{2} < 3$. 
In the A-type AFM state, the orbital PS angle is uniquely determined as 
$\theta=\cos^{-1} \{ 2J_{2}/(5J_{1}-J_{2}+6g)\}$.
By taking the MF results into account, for the following MC calculations, 
we choose the parameter set as 
$(J_{1}/J_{2}, g/J_{2})=(2.9, 5)$. 
In these values, 
the OO appears at much higher temperature than the N$\rm \acute e$el one, 
and the A-type AFM is realized near the phase boundary between FM and A-type AFM. 
These are suitable to demonstrate the magnetic structure change due to dilution. 
The MC simulation results in the realistic parameter set for LaMnO$_3$ will be introduced in the Sect.~\ref{sec:level6}. 
%
%The second set (2.55, 5 )is considered to be realistic values for LaMnO$_3$. 
%We compare the calculated $T_{\rm OO}$, N${\rm \acute e}$el temperature $T_{N}$, and 
%the spin stiffness constant in this parameter set with the experimental values. 
%\textcolor{red}
%{``Meaning of the following sentences is not clear"; 
%Since cos $\theta$ is very small for these two parameter sets, the PSs at ground states almost align the $T_{x}$ direction
%which is orthogonal to $T_{z}(\tau^{z})$ direction. 
%If the A-type antiferromagnetism staggered along $x,(y)$ direction,
%the PSs at ground states is orthogonal to $\tau^{x(y)}$ direction.
%}

\begin{figure}
\begin{center}
\includegraphics[width=0.8\columnwidth]{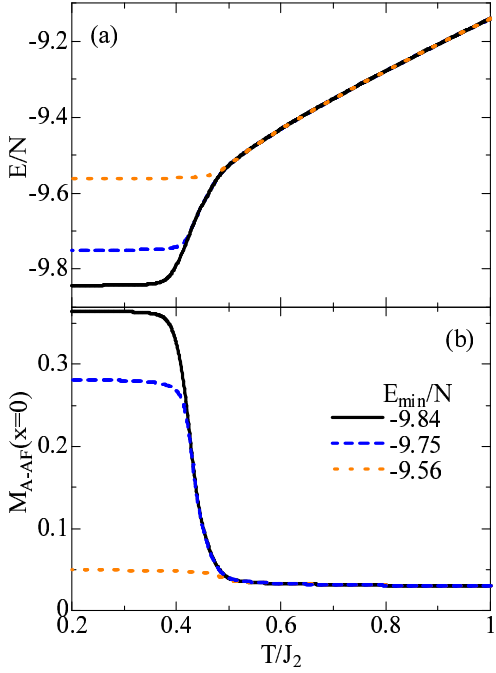}
%\scalebox{1.0}{\includegraphics[width=9cm,height=4.5cm,clip]{sogmaafemincomparison.eps}}
\end{center}
\caption{(a) Total energy $E$, and (b) the A-type AFM correlation function $M_{\rm A-AF}(x=0)$ calculated for several values of the minimum energy $E_{\rm min}$ in the WL method. 
System size is chosen to be $L=10$. }
\label{sogemin}
\end{figure}  
In the MC simulation, we utilize the WL method in $L \times L \times L$ site cluster 
($L=6-10$) with the periodic-boundary condition.
The spin operator $\bf{S}_{i}$ in the Hamiltonian is treated as a 3D 
classical vector with an amplitude of 1/2. 
In the simulation, 2$\times$10$^{7}$ MC steps are spent for measurement after calculating the histogram for the density of states. 
Physical quantities are averaged over 10MC samples at each parameter set.
We notice again the lowest energy edge  
$E_{\rm min}$ in the density of states which is introduced in Sec.~III.  
In Fig.~\ref{sogemin}, we show the $E_{\rm min}$ dependence of 
the total energy, $E$, and A-type AFM correlation function, $M_{\rm A-AF}(x)$ defined by 
\begin{align}
M_{\rm A-AF}(x)=\frac{1}{N(1-x)}
%\sqrt{ 
\left \langle 
\biggl \{ \sum_{i , l }(-1)^{i_{l}} \varepsilon_{i}\bf{S}_{i} \biggr \}^{2} \right \rangle^{1/2}
%},
\label{sog1}
\end{align}
where $i_{l}$ for $l=(x, y, z)$ represents the $l$ component of the coordinate at site $i$.
%In this definition, the maximum value of $M_{\rm A-AF}(x)$ is 0.5 for any value of $x$. 
%
The results in Fig.~\ref{sogemin}(a) imply that 
the temperature below which the total energy is flat is determined by an adopted value of $E_{\rm min}$. 
This temperature is denoted as $T_{\rm min}$ from now on. 
As shown in Fig.~\ref{sogemin}(b), in the case of $E_{\rm min}=-9.75$ $(-9.84)$, 
an obtained $M_{\rm A-AF}(x=0)$ below $T_{\rm min}$ is about $55\%$ $(75\%)$ of 
its maximum value of 1/2. 
That is, $T_{\rm min}$ at $E_{\rm min}=-9.84$ is lower than the N$\rm \acute e$el temperature of the A-type AFM. 
Although a saturated value of $M_{\rm A-AF}(x=0)$ is less than 0.5, this result is enough to examine the ordering temperature. 
We chose $E_{\rm min}=-9.84$ in the following MC simulation 
and focus on change of the magnetic ordering temperature due to dilution. 

\begin{figure}
\begin{center}
\includegraphics[width=0.8\columnwidth]{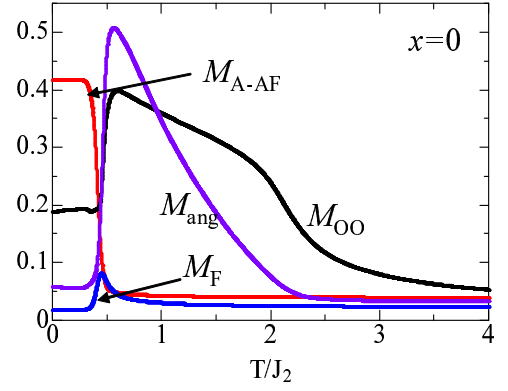}
%\scalebox{1.0}{\includegraphics[width=5.5cm,height=4.5cm,clip]{sogsaaffoaf.eps}}
\end{center}
\caption{Temperature dependence of the orbital correlation function $M_{\rm OO}(x=0)$, 
the PS angle function $M_{\rm ang}(x=0)$, the A-type AFM one $M_{\rm A-AF}(x=0)$, and the FM one $M_{\rm F}(x=0)$.
System size is chosen to be $L=8$.}
\label{sogoafsaafsf}
\end{figure}
\begin{figure}
\begin{center}
\includegraphics[width=0.8\columnwidth]{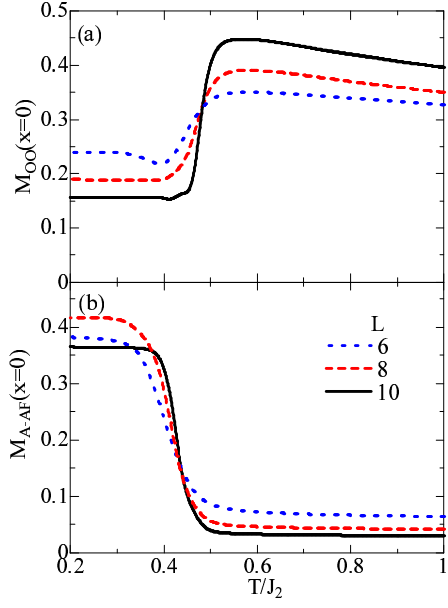}
%\scalebox{1.0}{\includegraphics[width=9.5cm,height=8cm,clip]{sogcomparison.eps}}
\end{center}
\caption{(a) System size dependence of the orbital correlation function 
$M_{\rm OO}(x=0)$, and (b) that of the A-type AFM correlation function $M_{\rm A-AF}(x=0)$. }
\label{sogoosaafoang}
\end{figure}

First, we show the results without impurities.
In Fig.~\ref{sogoafsaafsf}, calculated $M_{\rm OO}(x=0)$, $M_{\rm A-AF}(x=0)$, and the ferromagnetic correlation function defined by 
\begin{align}
M_{\rm F}(x)=\frac{1}{N(1-x)} 
%\sqrt{ 
\left \langle \biggl ( \sum_{i} \varepsilon_{i} \bf{S}_{i} \biggr )^{2} \right \rangle^{1/2}
%}, 
\label{sog2}
\end{align}
are presented. 
The staggered-type orbital correlation function $M_{\rm OO}(x=0)$ abruptly increases around $T/J_2=2.5$ which corresponds to the OO temperature $T_{\rm OO}(x=0)$. 
This value is consistent with the previous results obtained in the model Hamiltonian ${\cal H}_{T}$; the effective orbital interaction in the present Hamiltonian ${\cal H}_{ST}$ with paramagnetic state is 
$J_{\rm orb}=g+3J_{1}/4-J_{2}/4$ where ${\bf S}_i \cdot {\bf S}_j$ in ${\cal H}_{ST}$ is replaced by zero. 
The obtained $T_{\rm OO}(x=0)=2.5J_2$ corresponds to $0.3J_{\rm orb}$ in the present parameter set. 
This value is consistent with 
$T_{\rm OO}(x=0)=0.344J$ obtained in Sect.~IV [see Fig.~\ref{egorder}(a)]. 
In Fig.~\ref{sogoafsaafsf}, the angle correlation function $M_{\rm ang}(x=0)$ starts to increase at $T_{\rm OO}(x=0)$.  
With decreasing temperature, at around $T/J_2=0.5 [\equiv T_{\rm N}(x=0)]$, the second transition occurs. 
The orbital correlation function $M_{\rm OO}(x=0)$ decreases abruptly, and $M_{\rm A-AF}(x=0)$ grows up. 
The ferromagnetic correlation function $M_{\rm F}(x=0)$ shows a small hump structure around $T_{\rm N}(x=0)$.
That is, $T_{\rm N}(x=0)$ is the N$\rm \acute e$el temperature of A-type AFM. 
The PS angle correlation $M_{\rm ang}(x=0)$ decreases and almost becomes zero below $T_{\rm N}(x=0)$.
This result indicates that, due to the magnetic transition, the PS angle is changed into $(\theta_{A}, \theta_{B}) \sim (\pi/2, -\pi/2)$ which is consistent with the MF results. 
In Figs.~\ref{sogoosaafoang}, size dependences of $M_{\rm A-AF}(x=0)$, $M_{\rm OO}(x=0)$ are presented.
With increasing $L$, changes of $M_{\rm OO}(x=0)$ and $M_{\rm A-AF}(x=0)$ at $T_{\rm N}(x=0)$ become 
steep, although a saturated values of $M_{\rm A-AF}(x=0)$ is still less than 0.5 due to 
a finite value of $|E_{\rm min}-E_{GS}|$ as mentioned above. 

\begin{figure}
\begin{center}
\includegraphics[width=0.8\columnwidth]{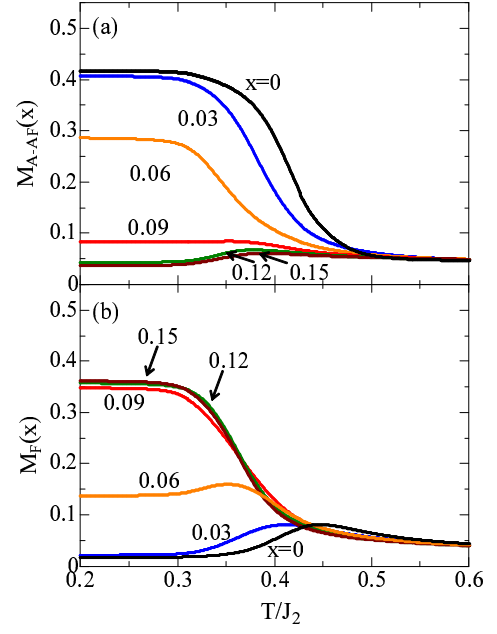}
%\scalebox{1.0}{\includegraphics[width=9cm,height=4.5cm,clip]{sogaaftoferr.eps}}
\end{center}
\caption{(a) Impurity concentration $x$ dependence of the A-type AFM correlation function $M_{\rm A-AF}(x)$, and (b) that of the FM correlation function $M_{\rm F}(x)$.
System size is chosen to be $L=8$.}
\label{sogaaftof}
\end{figure}  
Impurity concentration $x$ dependences of $M_{\rm A-AF}(x)$ and $M_{\rm F}(x)$ are presented in Fig.~\ref{sogaaftof}.
With increasing $x$ from the $x=0$ case, $M_{\rm A-AF}(x)$ decreases gradually and almost disappears around $x=0.09$. 
On the other hand, $M_{\rm F}(x)$, which shows a small hump structure around $T/J_2=0.4$ at $x=0$, increases with $x$, and takes about 0.35 in the case of $x>0.09$. 
That is to say, the magnetic structure is changed from A-AFM into FM by dilution. 
At $x=0.06$, both $M_{\rm A-AF}(x)$ and $M_{\rm F}(x)$ coexist down to the lowest temperature 
in the present simulation. 
This is supposed to be a cant-type magnetic order or a magnetic phase separation 
of the FM and A-type AFM phases.

\begin{figure}
\begin{center}
\includegraphics[width=0.9\columnwidth]{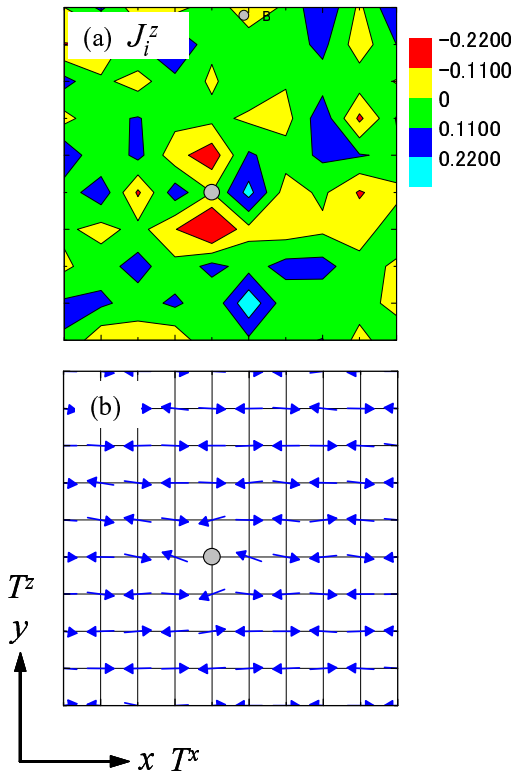}
%\scalebox{1.0}{\includegraphics[width=8cm,height=11cm,clip]{sogsnapshotb.eps}}
\end{center}
\caption{
(a) Contour map of $J_i^{z}$ defined in Eq.~(\ref{sog4}), and (b) a snapshot of the PS configuration around an impurity in the $xy$ plane obtained in the MC method. 
A filled circle represents an impurity. 
Temperature is chosen to be $T/J_2=0.3$. 
}
\label{sogsintorb}
\end{figure}
\begin{figure}[t]
\begin{center}
\includegraphics[width=0.7\columnwidth]{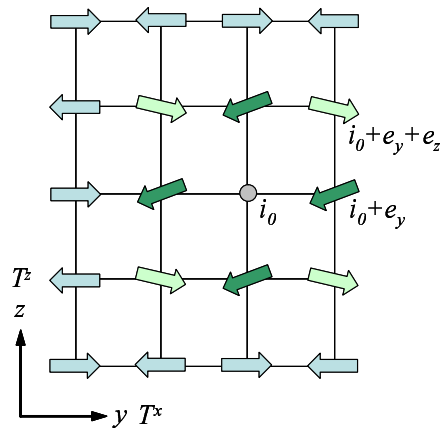}
%\scalebox{1.0}{\includegraphics[width=5cm,height=4cm,clip]{sogsichange.eps}}
\end{center}
\caption{
A schematic PS configuration around an impurity at site $i_0$.
A filled circle represents an impurity.}
\label{sogsintch}
\end{figure}  
To clarify the mechanism of the magnetic structure change due to dilution, the effective magnetic interaction and the PS configuration are examined. 
Here, the AFM stacking in the A-type AFM structure is chosen to be parallel to the $z$ axis. 
%In order to make clear a discussion, $H_{eff}$ is rewritten by
%\begin{align}
%H_{eff}&=\sum_{<ij>}n_{i}\varepsilon_j\bf{S}_{i} \cdot \bf{S}_{j} \{ 2(J_{1}+J_{2})\tau_{i}^{l}\tau_{j}^{l}+2J_{2}(\tau_{i}^{l}+%\tau_{j}^{l}) \\
% &+\frac{3}{2}J_{2}-\frac{1}{2}J_{1} \} +(2g+\frac{3}{2}J_{1}-\frac{1}{2}J_{2})\sum_{<ij>}n_{i}\varepsilon_j\tau_{i}^{l}\tau_{j}^%{l},
%\label{sog3}
%\end{align}
%where constant term is neglected.
%
The effective magnetic interaction $ J_{i}^l$ is defined 
such that the Hamiltonian ${\cal H}_{ST}$ in Eq.~(\ref{mod1}) is rewritten as 
${\cal H}_{ST}=\sum_{\langle ij \rangle} J_{i}^l \bf{S}_{i} \cdot \bf{S}_{j}$. 
The explicit form of the effective interaction is given as 
\begin{align}
J_{i}^z= 2 \left ( J_{1}+J_{2} \right ) T_{i}^z T_{j}^z
+2J_{2} \left ( T_{i}^{z}+T_{j}^{z} \right )+\frac{3}{2}J_{2}-\frac{1}{2}J_{1} ,  
\label{sog4}
\end{align}
where we consider a NN pair of sites $i$ and $j(=i+\hat e_z)$ along the $z$ direction, since we are interested in the magnetic structure along $z$. 
A contour map of the effective interaction $J_{i}^z$, and a snapshot of the PS configurations in the same $xy$ plane are presented in Fig.~\ref{sogsintorb}(a) and (b), respectively. 
Signs of $J_{i}^z$ in almost all region are positive (antiferromagnetic), reflecting the A-AFM structure. 
At the neighboring sites of the impurity along the $y$ direction, 
$J_{i}^z$s are negative (ferromagnetic). 
Away from the impurity, PS are ordered as  $\pm T^x$ in the staggered-type OO. 
Near the impurity, PS tilt from $\pm T^x$ and finite components of $T^z$ appears. 
This tilting of PS is seen in the results 
of ${\cal H}_T$ as explained in Sec.~\ref{sec:level4}. 
%
%We explaine mechanism of the ferromagnetic interaction induced by orbital dilution. 
Based on these numerical simulation, 
we explain mechanism of the magnetic structure change due to dilution. 
Start from the staggered-type orbital ordered state of $(T^x, T^z)=(\pm 1/2, 0)$. 
Introduce one impurity at a site $i_0$ 
which belongs to the $T^x=1/2$ sublattice, and focus on the PS configuration and the effective exchange interaction at sites $i_0+\hat e_m$ and $i_0+\hat e_m+\hat e_z$ for $m=(x,y)$
(see Fig.~\ref{sogsintch}).  
As explained in Sect.~\ref{sec:level4}, orbital dilution induces the PS tilting so as to gain 
the energies of the bonds where an impurity does not occupy. 
Thus, PS at site $i_0 + m\hat e$ tilts from $\theta=3\pi/2$ to $3\pi/2 + \delta \theta (- \delta \theta)$ for $m=x (y)$ with a positive angle $\delta \theta$.  
Since the orbital interaction is the staggered-type, 
the bilinear term $T_{i_0+ \hat{ e}_m}^z T_{i_0+ \hat{e}_m +\hat{e}_z}^z$ in Eq.~(\ref{sog4}) are negative 
for both the $m=x$ and $y$ cases. 
As for the linear term in Eq.~(\ref{sog4}), 
there is a relation 
$(T_{i_0+\hat e_x}^z+T_{i+\hat e_x+\hat e_z}^z) =-(T_{i_0+\hat e_y}^z+T_{i+\hat e_y+\hat e_z}^z) $. 
That is, contribution of this linear term to the spin alignment along $z$, which is determined by a sum of  $J_{i_0+\hat e_x}^z$ and $J_{i_0+\hat e_y}^z$,  
is canceled out.  
Therefore, 
when the first term in Eq.~(\ref{sog4}) overcomes the positive constant $3J_2/2-J_1/2$, 
$J^z_i$ becomes negative and the ferromagnetic alignment along the $z$ direction 
is stable around the impurity sites.

\section{\label{sec:level6}Summary and discussion}

In this section, we discuss implications of the present numerical calculations 
on the recent experimental results in the transition-metal compounds. 
First we have remarks on the relation between the calculated results of ${\cal H}_T$ shown in Sect.~\ref{sec:level4} and the experiments in KCu$_{1-x}$Zn$_x$F$_3$.~\cite{tatami07}
As shown in Sect~\ref{sec:level4}, $T_{\rm OO}(x)$ rapidly decreases with increasing $x$ in comparison with dilute magnets (see Fig.~\ref{egphase}). 
Although the critical concentration ($x \sim 0.2-0.5$), where the OO disappears, depends on the calculation methods, that is, MC and CE, these values are far below the percolation threshold ($x_p=0.69$). 
This result is consistent qualitatively with the Zn concentration dependence of the OO temperature in KCu$_{1-x}$Zn$_x$F$_3$ where OO vanishes around $x=0.45$. 
One of the discrepancies between the theory and the experiments are seen in their quantitative values of the critical impurity concentration where OO disappears. 
Some of the reasons of this discrepancy may be attributed to the anharmonic JT coupling and the long-range PS interactions due to the spring constants beyond the NN ions and so on, both of which are not taken into account in the present calculation. 
The former subject, i.e. the anharmonic JT coupling, induces the anisotropy in a bottom of the adiabatic potential of the $Q_x-Q_z$ plane, and prevents the PS tilting around impurity sites. 
This effect on the reduction of $T_{\rm OO}(x)$ was studied briefly in Ref.~\onlinecite{tanaka05}. 
It was shown that, in the realistic parameter values, the reduction of $T_{\rm OO}(x)$ becomes moderate by the anharmonic coupling, but it is still steeper than that in dilute magnets. 
Another factor which may explains the discrepancy between the theory and the experiments is the quantum aspect for the orbital degree of freedom. 
In the results obtained by the quantum CE method as shown in Fig.~\ref{egphase}, the critical $x$ for $T_{\rm OO}(x)=0$ is larger than the results by other two classical calculations for the orbital model and is close to the experimental value of $x=0.45$. 
This may be due to the fact that quantum fluctuation in low temperatures weakens the low dimensional character in the OO state and prevents a collapse of OO against dilution. 
This kind of quantum effects in the dilute orbital system was examined by the present authors in the two dimensional quantum orbital model.~\cite{tanaka07} 
It was shown that the reduction of $T_{\rm OO}(x)$ due to dilution is weaker than that in the classical orbital model. 

We briefly mention the orbital PS tilting due to dilution. 
Similar phenomena are known as a quadrupolar glass state in molecular crystals where 
different kind interactions between molecules with quadruple moment coexists.~\cite{binder92}
A kind of glass state in terms of the quadrupole moment appears with increasing 
randomness for the interactions. 
We suggest a possibility that the present observed PS tilting accompanied with the lattice distortion 
of ligand ions is able to be detected experimentally. 
One of the most adequate experimental techniques are the pair-distribution function method  
by the neutron diffraction experiments, and X-ray absorption fine structure (XAFS) where the incident x-ray energy is tuned at the absorption edge of the impurity ions. 
This observation may work as a check for the present scenario in the dilute orbital system. 

\begin{figure}
\begin{center}
\includegraphics[width=0.7\columnwidth]{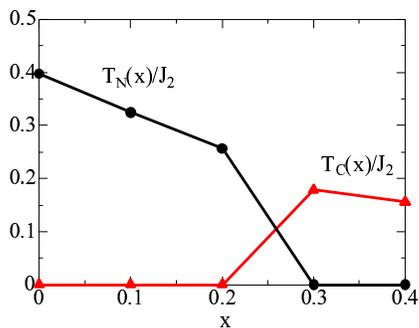}
%\scalebox{1.0}{\includegraphics[width=9cm,height=4.5cm,clip]{tnc.eps}}
\end{center}
\caption{
Impurity concentration dependence of the A-type AFM transition temperature, $T_{\rm N}(x)$, and 
the FM one, $T_{\rm c}(x)$, 
calculated in the realistic parameter values for LaMn$_{1-x}$Ga$_x$O$_3$. 
The parameters and the system size are chosen to be $(J_{1}/J_{2}, g/J_2)=(2.5, 5)$ 
and $L=8$, respectively.}
\label{fig:tctn}
\end{figure}
%
%\begin{figure}
%\begin{center}
%\scalebox{1.0}{\includegraphics[width=9cm,height=4.5cm,clip]{sogaaff525.eps}}
%\end{center}
%\caption{(a) the temperature dependence of $M_{AAF}(x)$. (b) the temperature dependence of $M_{F}(x)$.}
%\label{sogmc525}
%\end{figure}
Next we discuss implications of the calculated results in Sect.~\ref{sec:level5} to the experimental results in LaMn$_{1-x}$Ga$_x$O$_3$.~\cite{farrell04,blasco02,sanches04,goodenough61,zhou01,zhou03}
By analyzing the spin-orbital coupled Hamiltonian ${\cal H}_{ST}$, we find that the magnetic structure is changed from the A-AFM order into the FM one. 
This calculation qualitatively explains the experimental results in LaMn$_{1-x}$Ga$_x$O$_3$ from the macroscopic point of view. 
In Sect.~\ref{sec:level5}, the parameter set is chosen to be close to the values for the A-type AFM/FM phase boundary, in order to demonstrate clearly the magnetic structure change due to the orbital dilution. 
Here we briefly introduce the numerical results obtained in the realistic parameter values. 
To evaluate the realistic values, we calculate the OO temperature, the N$\rm \acute e$el temperature by the MF approximation, and the spin wave stiffness by the spin wave approximation from ${\cal H}_{ST}$, and compare the experimental results in  LaMnO$_3$. 
Then, we set up the parameters as $(J_{1}/J_{2}, g/J_2)=(2.5, 5)$. 
The $x$ dependences of the magnetic transition temperatures are presented in Fig.~\ref{fig:tctn}. 
With increasing $x$ from $x=0$, $T_{\rm N}(x)$ of the A-type AFM order gradually decreases, and around $x=0.2$, the A-type AFM is changed into the FM order which remains at least to $x=0.4$. 
%Around $x=0.2$, the two magnetic correlation functions, $M_{\rm A-AF}(x)$ and $M_{\rm F}(x)$, coexist, 
%as seen in Fig.~\ref{sogaaftof}. 
In semi-quantitative sense, this result is consistent with the experimental magnetic phase diagram in LaMn$_{1-x}$Ga$_x$O$_3$. 
However, one of the discrepancies is that the canted phase survives up to $x=0.4$ in LaMn$_{1-x}$Ga$_x$O$_3$. 
This difference between the theory and the experiments 
is supposed to be due to the $t_{2g}$ spins in Mn sites and the antiferromagnetic superexchange interaction between them which are not included explicitly in the present calculation. 
This interaction stabilizes the A-type AFM phase in comparison with the FM one, and maintains the canted phase up to a higher $x$ region. 

In summary, we present a microscopic theory of dilution effects in the $e_g$ orbital degenerate system. 
We analyze the dilution effects in the $e_g$-orbital Hamiltonian without spin degree of freedom, ${\cal H}_T$, and the spin and $e_g$ orbital coupled Hamiltonian, ${\cal H}_{ST}$. 
The classical MC simulation and the CE method are utilized. 
It is shown that the OO temperature decreases rapidly with increasing dilution. 
From the system size dependence of the orbital correlation function in the MC method, the OO is not realized at the impurity concentration $x=0.2$. 
Tilting of orbital PS around impurity is responsible for this characteristic reduction of $T_{\rm OO}(x)$. 
This is consequence of the bond dependent interaction between the inter-site orbitals. 
In the analyses of the spin-orbital coupled model, the magnetic structure is changed from the A-type AFM structure into the FM one by dilution. 
This is explained by changing of the magnetic interaction due to the orbital PS tilting around the impurity. 
The present results explain microscopically the novel dilution effects in KCu$_{1-x}$Zn$_{x}$F$_{3}$ and LaMn$_{1-x}$Ga$_{x}$O$_{3}$, and provide a unified picture for the dilution effect in the orbital ordered system. 

\begin{acknowledgments}
The authors would like to thank 
Y.~Murakami, M.~Matsumoto, and H.~Matsueda for their valuable discussions. 
The authors also thank  T.~Watanabe and J.~Nasu for their critical reading of the manuscript. 
This work was supported by JSPS KAKENHI (16104005), and 
TOKUTEI (18044001, 19052001, 19014003) from MEXT, 
NAREGI, and CREST.
One of the authors (T.T,) thanks the financial support from JSPS. 
\end{acknowledgments}

%\appendix

%\section{Appendixes}
%\begin{verbatim}
%\appendix
%\section{}
%\end{verbatim}
%will produce an appendix heading that says ``APPENDIX A'' and
%\begin{verbatim}
%\appendix
%\section{Background}
%\end{verbatim}
%will produce an appendix heading that says ``APPENDIX A: BACKGROUND''
%(note that the colon is set automatically).

%If there is only one appendix, then the letter ``A'' should not
%appear. This is suppressed by using the star version of the appendix
%command (\verb+\appendix*+ in the place of \verb+\appendix+).

%\bibitem{aa}aa
%

$^{\ast}$ Present address: 
The Institute for Solid State Physics, University of Tokyo, Kashiwa, Chiba 277-8581, Japan. 

\end{document}